\documentclass[12pt]{article}
\usepackage{epsfig,graphicx}
\usepackage{ch2005}

\def\beq{\begin{equation}}
\def\eeq#1{\label{#1}\end{equation}}
\def\eeqn{\end{equation}}

\def\beqa{\begin{eqnarray}}
\def\eeqa#1{\label{#1}\end{eqnarray}}
\def\eeqan{\end{eqnarray}}

\def\Dslash{\not{\hbox{\kern-4pt $D$}}}
\def\dslash{\not{\hbox{\kern-2pt $\del$}}}

\newcommand{\tev}{\ensuremath{\mathrm{\,Te\kern -0.1em V}}\xspace}
\newcommand{\gev}{\ensuremath{\mathrm{\,Ge\kern -0.1em V}}\xspace}
\newcommand{\mev}{\ensuremath{\mathrm{\,Me\kern -0.1em V}}\xspace}
\newcommand{\kev}{\ensuremath{\mathrm{\,ke\kern -0.1em V}}\xspace}
\newcommand{\ev}{\ensuremath{\mathrm{\,e\kern -0.1em V}}\xspace}
\newcommand{\gevc}{\ensuremath{{\mathrm{\,Ge\kern -0.1em V\!/}c}}\xspace}
\newcommand{\mevc}{\ensuremath{{\mathrm{\,Me\kern -0.1em V\!/}c}}\xspace}
\newcommand{\gevcc}{\ensuremath{{\mathrm{\,Ge\kern -0.1em V\!/}c^2}}\xspace}
\newcommand{\mevcc}{\ensuremath{{\mathrm{\,Me\kern -0.1em V\!/}c^2}}\xspace}

\def\mus  {\ensuremath{\rm \,\mus}\xspace}

\def\mus        {\ensuremath{\,\mu{\rm s}}\xspace}    

\begin{document}


\Title{Microquasar hadronic jets at very high-energy gamma-rays}
\bigskip


%
\label{Bosch-RamonStart}

%
\author{Valent\'i Bosch-Ramon(1)\index{Bosch-Ramon, V.}, Felix A. 
Aharonian(2)\index{Aharonian, F. A.} and Josep M. Paredes\index{Paredes, J.
M.}(1)}

%
\address{(1)Departament d'Astronomia i Meteorologia, Universitat de Barcelona\\
Av. Diagonal 647, E-08028 Barcelona, Catalonia, Spain\\
(2)Max-Planck-Institut fur Kernphysik\\
Saupfercheckweg 1, Heidelberg, 69117, Germany}

\makeauthor\abstracts{
Microquasars (MQs) present emission over the whole spectrum, from radio wavelengths
to gamma-rays. The microquasar spectral energy distribution is very complex,
being a signature of the different physical processes that generate the
radiation emitted by these objects. In this work, we estimate the amount of 
broad-band emission produced by relativistic protons, released from the jet 
of a MQ, interacting with high density regions of the ISM. We show that a 
two components source, the
microquasar itself and the region of interaction between the jets and the ISM,
could be unveiled by the new instruments at high-energy and very high-energy
gamma-rays.
}

\section{Introduction} 

The likely association between microquasars and gamma-ray sources
(\cite{Paredes-ch2005}), altogether with the evidence of ion presence in MQ jets
(\cite{Migliari-ch2005}), might point to the possibility that these objects could
produce variable and point-like as well as steady and perhaps extended gamma-ray
emission. On the one hand, the radiation coming from the microquasar itself,
particularly from its jets (likely variable and point-like), could reach very high
energies and significant fluxes.  The emission produced in the jets of MQs has been
studied elsewhere, and leptonic models (e.g. \cite{Bosch-Ramona-ch2005}) and hadronic
models (e.g. \cite{Romero-ch2005}) have been proposed in order to explain the likely
association between microquasars and gamma-ray sources. On the other hand, accelerated
protons in a jet termination shock could diffuse through the ISM generating detectable
amounts of low energy as well as high energy emission (steady and perhaps
extended) when these protons reach higher density regions (i.e. clouds). 

\begin{table}[htbp]
\begin{center}
\begin{tabular}{|c|c|}
\hline
Parameter & Value \\ 
\hline
diffusion coefficient normalization constant at 10 GeV & $10^{27}$~cm$^2$~s$^{-1}$ \\
diffusion power-law index & 0.5 \\
ISM density & 0.1~cm$^{-3}$ \\
cloud density & 10$^4$~cm$^{-3}$ \\
mass of the high density region/cloud & $3\times10^4~M_{\odot}$ \\
magnetic field within the cloud & $5\times10^{-4}$~G \\
IR radiation energy density within the cloud & 10~eV~cm$^{-3}$ \\
power-law index of the high energy protons & $\sim 2$ \\
maximum energy of the high energy protons & $\sim 10^5$~GeV \\
kinetic luminosity of accelerated protons in the MQ jet & 10$^{37}$~erg~s$^{-1}$ \\
kinetic energy of accelerated protons in the impulsive ejection & 10$^{48}$~erg \\
distance between the MQ and the cloud & 10~pc \\
\hline
\end{tabular}
\caption{Adopted parameter values} 
\label{tab:Bosch-Ramon-page}
\end{center}
\end{table}

\section{MQ-cloud interactions and gamma-ray production}

Jets of MQs should end somewhere, although it is still unclear the way how they
terminate (\cite{Heinza-ch2005}, \cite{Heinzb-ch2005}). Assuming that the jet
has an important population of protons and such protons are accelerated in the
terminal part of the jet (getting a power-law energy distribution), interactions
between those high energy particles and nearby clouds can lead to the creation of
neutral and charged pions. Then, neutral pions will decay to gamma-rays photons while charged
pions will decay to e$^-$ and e$^+$. These secondary particles can produce
significant levels of synchrotron (from radio frequencies to X-rays) and
Bremsstrählung emission (from soft gamma-rays to the TeV range), 
and generally with much
less efficiency, inverse Compton high energy 
emission through interaction with the ambient
infrared photons. For further details, see \cite{Bosch-Ramonb-ch2005}.

\begin{figure}[htb]
\begin{center}
\epsfig{file=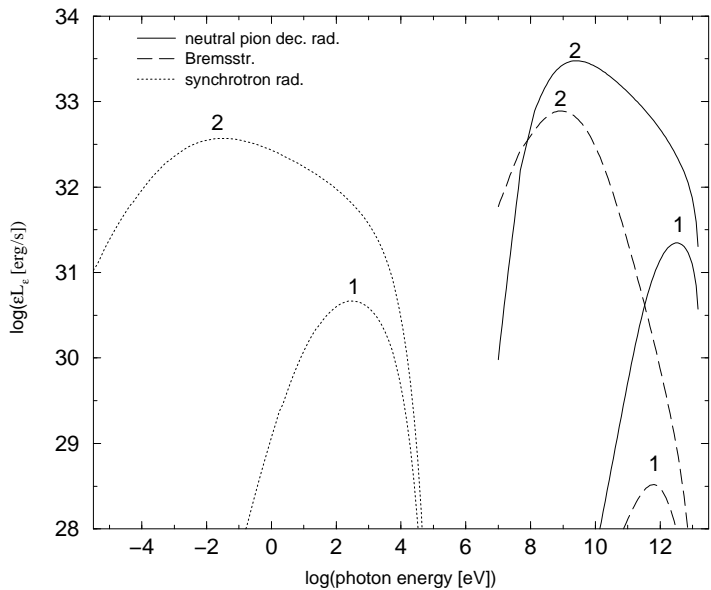,height=80mm}
\caption{SED for a continuous MQ from radio to vcry high-energy gamma-rays at two
different ages: t=100 yr (1), t=10000 yr (2). Note that the source could be
detected almost at all the frequencies if located at galactic distances.}
\label{fig:Bosch-Ramon-fig1}
\end{center}
\end{figure}

\begin{figure}[htb]
\begin{center}
\epsfig{file=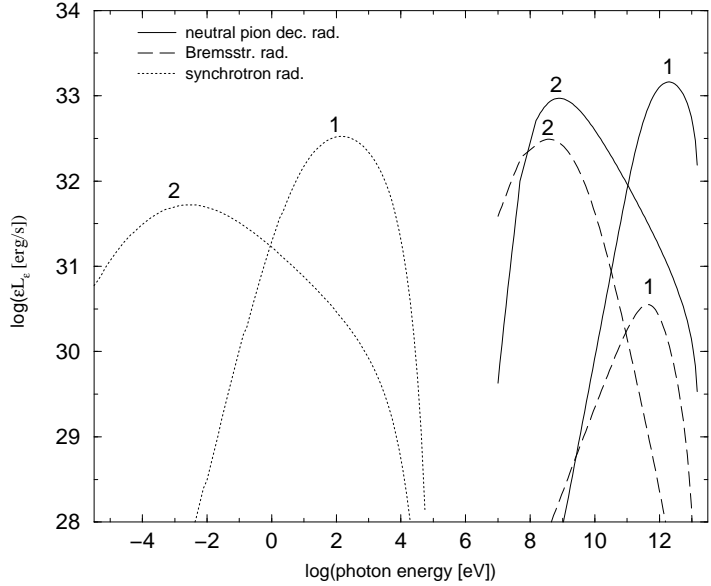,height=80mm}
\caption{The same as in Fig. 1, but for an impulsive microquasar. Note that the
detection would be more probable for short times after the impulsive event.}
\label{fig:Bosch-Ramon-fig2}
\end{center}
\end{figure}

Protons diffuse through the ISM, and the diffusion coefficient has been assumed to
be a power-law in energy. Due to propagation effects, the outcomes of the
interactions between the protons released from the jet and protons in the clouds can differ
strongly depending on the age, the nature (impulsive or continuous) of the
accelerator and the distance between this and the cloud (see
\cite{Aharonian-ch2005}). In this work we study both cases, the continuous and the
impulsive one, considering as a target a nearby cloud at several pc of distance. In
Table~\ref{tab:Bosch-Ramon-page}, the adopted parameter values are presented. We note that in the gamma-ray
band, due to the characteristics of the proton-proton interaction, the main energy
release channel is the neutral pion-decay, being dominant over the Bremsstr\"ahlung
component. In Figs.~\ref{fig:Bosch-Ramon-fig1} and \ref{fig:Bosch-Ramon-fig2}, the
broad-band spectral energy distributions (SEDs) for the impulsive and the continuous
case are shown. Note the different slopes and fluxes depending on the case. 
For a continuous MQ, the spectrum reaches higher luminosities ,and gets softer, 
when the source is "old" and gets the steady regime. For a continuous MQ, The 
higher luminosities, with a hard spectrum, can be observed when the source is "young".

Although the number of known MQs is reduced at the present moment (about 16, see
\cite{Ribo-ch2005}, \cite{Paredesb-ch2005}), they could be a significant fraction of the unidentified
gamma-ray sources in the galactic plane (\cite{Bosch-Ramona-ch2005}). If this
association as well as their intrinsic nature of gamma-ray emitters are confirmed,
these objects could turn out to be complex sources in the gamma-ray sky with
compounded spatial and spectral properties at very high energies. Our bet is that,
if observed with reasonable exposure times at high-energy and very high-energy
gamma-rays, MQs will be eventually detected. Otherwise, the existence of steady and
perhaps extended gamma-ray radiation surrounding a variable point-like gamma-ray
source could be a strong hint of the presence of a MQ interacting with its
environment and, in such a case, multiwavelength observations would be unavoidable
to disentangle the real nature and connections between both components, and
broad-band spectrum models like the one presented here can help to understand them.

\section{Summary}

The study of the emission coming from high density regions of the ISM around
MQs can give us information not only about such regions, but also about the
objects that are injecting energy in those regions through high energy protons.
In this work, we explored the observational implications at very high energies 
of the presence of a MQ close to a cloud.
New gamma-ray instruments like the ground-based Cherenkov telescopes or the
next generation of gamma-ray satellites could be able to detect separately the
emission coming from a MQ and the region around, if significant hadronic
interactions take place there. 

\section*{Acknowledgments}
V.B-R. and J.M.P. acknowledge partial support by DGI of the spanish Ministerio
de Educación y Ciencia (former Ministerio de Ciencia y Tecnolog\'ia) under
grant AYA-2001-3092 and AYA2004-07171-C02-01, as well as additional support
from the European Regional Development Fund (ERDF/FEDER). During this work,
V.B-R. has been supported by the DGI of the spanish Ministerio de Educaci\'on y
Ciencia under the fellowship BES-2002-2699.

%
\label{Bosch-RamonEnd}
 
\end{document}